\documentclass[runningheads]{llncs}
\usepackage[T1]{fontenc}
\usepackage{graphicx}
\usepackage{amsmath}
\usepackage{amssymb}
\usepackage{bm} 
\def \vect#1{\bm{#1}}
\def \matr#1{\bm{#1}}

\begin{document}
\title{Data-Assimilated Model-Based Reinforcement Learning for Partially Observed Chaotic Flows}
\titlerunning{Data-Assimilated Model-Based Reinforcement Learning}
%
\author{Defne E. Ozan \inst{1}\orcidID{0009-0006-4208-9981} \and Andrea N{\'o}voa \inst{1} \orcidID{0000-0003-0597-8326} \and Luca Magri \inst{1,2,3} \orcidID{0000-0002-0657-2611} }
\authorrunning{D.E. Ozan, A. N{\'o}voa, L. Magri}
%
\institute{
Imperial College London, Department of Aeronautics, Exhibition Road, London SW7 2BX, UK \\
\email{d.ozan@imperial.ac.uk, a.novoa@imperial.ac.uk, l.magri@imperial.ac.uk}\\
\and
The Alan Turing Institute, London NW1 2DB, UK \and
Politecnico di Torino, DIMEAS, Corso Duca degli Abruzzi, 24 10129 Torino, Italy
}
\maketitle              
\begin{abstract}
The goal of many applications in energy and transport sectors is to control turbulent flows. However, because of chaotic dynamics and high dimensionality, the control of turbulent flows is exceedingly difficult. Model-free reinforcement learning (RL) methods can discover optimal control policies by interacting with the environment, but they require full state information, which is often unavailable in experimental settings. We propose a data-assimilated model-based RL (DA-MBRL) framework for systems with partial observability and noisy measurements. Our framework employs a control-aware Echo State Network for data-driven prediction of the dynamics, and integrates data assimilation with an Ensemble Kalman Filter for real-time state estimation. An off-policy actor-critic algorithm is employed to learn optimal control strategies from state estimates. The framework is tested on the Kuramoto-Sivashinsky equation, demonstrating its effectiveness in stabilizing a spatiotemporally chaotic flow from noisy and partial measurements.
\end{abstract}
\section{Introduction}{\label{sec:introduction}}
Turbulence arises in many engineering applications involving energy and transportation and has implications for efficiency and energy consumption of the system of interest. Closed-loop control has been key for taming turbulence for drag reduction of transport vehicles, lift increase of airfoils, efficiency increase of wind turbines, and mixing enhancement of combustors~\cite{brunton2015ClosedLoopTurbulenceControla}. While these control methods are well established for linear systems, finding effective and robust control strategies in turbulent systems is challenging due to the nonlinear and chaotic nature of the dynamics. Furthermore, turbulence spans over multiple spatial and temporal scales, requiring high-dimensional state spaces and expensive computations, which makes it difficult for control to work in real time. These challenges have motivated the development of data-driven control methods. Reinforcement learning (RL) has been imported in the field of flow control because it offers a model-free framework that can learn optimal control policies directly from data. Recent advancements in deep RL have demonstrated success in learning policies for flow control tasks such as drag reduction and separation control~e.g.,~\cite{rabault2019ArtificialNeuralNetworks}. However, because RL methods are typically Markov Decision Processes, they require access to the full state information, i.e., full observability. This poses a limitation in applications in which only a subset of the full state can be measured as they are Partially Observed Markov Decision Processes~\cite{krishnamurthy2016PartiallyObservedMarkov}. Previous approaches have addressed partial observability by incorporating a memory of the system’s measurements~\cite{xia2024ActiveFlowControl}, recurrent or attention-based architectures~\cite{weissenbacher2025ReinforcementLearningChaotic}. 
Data assimilation methods, specifically sequential assimilation methods, are commonly used to enable real-time predictions in numerical weather forecasting and fluid dynamics in partially-observed systems~\cite{evensen2009DataAssimilationEnsemblea}. In this paper, we employ sequential data assimilation  to integrate noisy and sparse measurements with a predictive model in order to estimate the full state of a controlled system. 
We develop an RL framework for chaotic flow control with the following objectives: (i) learn a data-driven model of the actuated dynamics of a chaotic flow, 
(ii) estimate the full state of the system in real time from partial, noisy observations using data assimilation, and 
(iii) learn an optimal control policy from the estimated state. 
This paper is organized as follows: Sec.~\ref{sec:problem_formulation} sets the control problem, Sec.~\ref{sec:methodology} presents the proposed framework, Sec.~\ref{sec:results} demonstrates it on the Kuramoto-Sivashinsky (KS) equation, and Sec.~\ref{sec:conclusions} concludes the paper.
\section{Problem formulation}\label{sec:problem_formulation}
We consider a deterministic discrete-time dynamical system of the form
\begin{equation}\label{eq:dyn_sys}
    \vect{s}(t+1) = \vect{\mathcal{F}}(\vect{s}(t), \vect{a}(t)), 
\end{equation}
where $\vect{s}(t) \in \mathbb{R}^{n_s}$ is the system's state, $\vect{a}(t) \in \mathbb{R}^{n_a}$ is the control action and $\vect{\mathcal{F}}: \mathbb{R}^{n_s} \times \mathbb{R}^{n_a} \rightarrow \mathbb{R}^{n_s}$ is the operator that governs the evolution of the system's dynamics in time. We focus on systems with chaotic dynamics, which are extremely sensitive to initial conditions making the system unpredictable in long time intervals. The system is Markovian, meaning that $\vect{s}(t)$ and $\vect{\mathcal{F}}$ encapsulate all necessary information to determine the next state, $\vect{s}(t+1)$. Our goal is to find the optimal set of actions, $\vect{a}^*(t)$ to optimize a certain objective functional constrained by the system's dynamics~\eqref{eq:dyn_sys}. In RL, this objective is formulated through a reward function, $r(\vect{s}, \vect{a}) \in \mathbb{R}$, which evaluates the immediate effect of an action. The agent, i.e., the controller, selects actions based on a state-feedback policy, i.e., $\vect{a}(t) = \vect{\pi}(\vect{s}(t))$. The value function
\begin{equation} 
    V^\pi(\vect{s}) = \mathbb{E}\left[ \sum_{t=0}^{\infty} \gamma^t r(\vect{s}(t), \vect{a}(t)) \mid \vect{s}(0) = \vect{s} \right]
\end{equation}
quantifies the expected return under policy $\vect{\pi}$, with discount factor $\gamma \in [0,1]$, which weighs the importance of future states. The agent is trained to learn the optimal policy $\vect{\pi}^*$ that maximizes the associated value function.
In practical applications, we acquire observations of the system's state, which are noisy and incomplete. We model the observation process as
\begin{equation} 
    \vect{o}(t) = \vect{\mathcal{M}}(\vect{s}(t)) + \vect{\epsilon}(t),
\end{equation}
where $\vect{o}(t) \in \mathbb{R}^{n_o}$ is the observation, $\vect{\epsilon}(t) \in \mathbb{R}^{n_o}$ is a zero-mean measurement noise, and $\vect{\mathcal{M}}: \mathbb{R}^{n_s} \rightarrow \mathbb{R}^{n_o}$ is the observation operator.
\section{Methodology}\label{sec:methodology}
To tackle control from partial observations, we propose a framework, data-assimilated model-based reinforcement learning (DA-MBRL), that has three components: (i) a predictive model of the system's dynamics, (ii) an ensemble-based data assimilation method for real-time state estimation, and (iii) an off-policy actor-critic RL algorithm for learning the control policy. If the system's equations are unknown or are too expensive for real-time simulation, a data-driven model can be utilized to learn and forecast the dynamics. For this purpose, we propose a control-aware formulation of the Echo State Network (ESN)~\cite{racca2023ControlawareEchoState}. The ESN~\cite{jaeger2004HarnessingNonlinearityPredicting} is a type of recurrent neural network (RNN) that is computationally cheap to train. To model actuated dynamics, we extend the ESN by incorporating control actions into the reservoir state update
\begin{equation}\label{eq:esn} 
    \vect{h}(t+1) = (1-\alpha)\vect{h}(t) + \alpha\tanh\big(\matr{W}_{in}[(\vect{y}_{in}(t)-\bar{\vect{y}})\odot\vect{g}_y; \vect{a}(t)]+\matr{W}\vect{h}(t)\big), 
\end{equation}
where $\vect{y}_{in}(t) \in \mathbb{R}^{n_y}$ is the input vector, $\vect{h}(t) \in \mathbb{R}^{n_h}$ is the reservoir state, $\matr{W}_{in} \in \mathbb{R}^{n_h \times (n_y + n_a)}$ is the input matrix, and $\matr{W} \in \mathbb{R}^{n_h \times n_h}$ is the state matrix. The input vector is standardized by the mean $\bar{\vect{y}} \in \mathbb{R}^{n_y}$ and the scaling $\vect{g}_y \in \mathbb{R}^{n_y}$, $g_{y,i} = 1/\sigma_{y,i}$ with $\sigma_{y,i}$ being the standard deviation of each $y_i, \; i = 1,2, \dots, n_y$. The time series prediction of the dynamics is enabled by the readout, which is defined as
\begin{equation}\label{eq:readout}
    \vect{\hat{y}}(t+1) = \matr{W}_{out}[\tilde{\vect{h}}(t+1); 1], 
    \;
    \tilde{h}_k(t) = 
    \begin{cases}
        h_k(t) & \text{if }k\text{ is odd}, \\
        h_k^2(t) & \text{if }k\text{ is even},
    \end{cases}
    \;
    k = 1,2, \dots, n_h,
\end{equation}
where $\vect{y} \in \mathbb{R}^{n_y}$ is the target output vector, $\vect{\hat{y}} \in \mathbb{R}^{n_y}$ is the predicted output vector, and $\matr{W}_{out} \in \mathbb{R}^{n_y \times (n_h+1)}$ is the output matrix. We include the squared reservoir state variables in the readout to break the antisymmetry of $\tanh$ and enable closed-loop prediction. The output weights $\matr{W}_{out}$ are trained in open-loop configuration via ridge regression, whilst the input and state matrices $\matr{W}_{in}$ and $\matr{W}$ are sparse, randomly generated, and not trained. The input matrix is obtained by first generating a matrix $\tilde{\matr{W}}_{in}$ such that each row contains only one nonzero element drawn from a uniform distribution $\sim \mathcal{U}(-1,1)$. The inputs are then scaled as $\matr{W}_{in} = [\xi_{in}\tilde{\matr{W}}_{in}^{y} \quad \xi_{a}\tilde{\matr{W}}_{in}^{a}]$, where $\tilde{\matr{W}}_{in}^y \in \mathbb{R}^{n_h \times n_y}$ and $\tilde{\matr{W}}_{in}^a \in \mathbb{R}^{n_h \times n_a}$ are the weights associated with the input and action vectors, respectively. The state matrix $\matr{W}$ is scaled as $\matr{W} = \rho \tilde{\matr{W}}$, where $\tilde{\matr{W}}$ is an Erd\H{o}s-Renyi matrix with a given average number of nonzero elements per row (connectivity, $n_{conn}$) drawn from a uniform distribution $\sim \mathcal{U}(-1,1)$ scaled to have unity spectral radius, so that $\rho$ defines the spectral radius of $\matr{W}$. The details of optimal training, hyperparameter selection and validation of ESNs can be found in~\cite{racca2021RobustOptimizationValidation}. In this paper, we train the ESN on the full state data from the system generated by random actuations, i.e., $\vect{y}_{in}(t) = \vect{y}(t) = \vect{s}(t)$. After training, the closed-loop prediction is enabled by setting $\vect{y}_{in}(t) = \vect{\hat{y}}(t)$, which means that in this configuration the ESN's state at time $t$ is described by the reservoir state~$\vect{h}(t)$ only. Due to the chaotic nature of the system, the ESN's prediction will decorrelate from the true state after some time steps. To correct the reservoir state, we employ the Ensemble Kalman Filter (EnKF)~\cite{evensen2009DataAssimilationEnsemblea}, which extends the Kalman Filter to nonlinear systems by approximating the covariance matrix of the state via an ensemble of state vectors, $\vect{h}_j, \; j = 1,2,\dots,m$. The EnKF consists of two steps: (i) the forecast step, denoted by $(\cdot)^f$, which propagates the ensemble of state vectors forward in time using~\eqref{eq:esn}, and (ii) the analysis step, denoted by $(\cdot)^a$, which updates the ensemble when observations become available. To account for nonlinear observation operators, we use the augmented formulation from~\cite{novoa2024InferringUnknownUnknowns}, $\vect{\psi}^f_j = [\vect{h}^f_j; \vect{\mathcal{M}}(\hat{\vect{s}}^f_j)]$ and define a new linear observation operator $\matr{M} = [\matr{0} \; \matr{I}]$, where $\matr{I}$ is the identity matrix. The analysis step is given by
\begin{equation}
    \vect{\psi}^a_j= \vect{\psi}^f_j + \matr{K}(\vect{o}_j - \matr{M}\vect{\psi}^f_j),
\end{equation}
where $\matr{K}$ is the Kalman gain $\matr{K} = \matr{C}^f_{\psi\psi}\matr{M}^\top(\matr{C}_{\epsilon\epsilon}+\matr{M}\matr{C}^f_{\psi\psi}\matr{M}^\top)^{-1}$, $\matr{C}^f_{\psi\psi}$ and $\matr{C}_{\epsilon\epsilon}$ denote the covariance matrices of $\vect{\psi}^f_j$ and $\vect{\epsilon}$, respectively. Under Gaussian assumptions, the ensemble approximation of the maximum a posteriori estimate of $\vect{h}(t)$ is the mean of the ensemble of $\vect{h}^a_j(t)$. 
By using the readout~\eqref{eq:readout} as a map from the reservoir state $\vect{h}$ to the system's true state $\vect{s}$, we compute the estimate of the system's full state from the mean of the ensemble of reservoir state vectors, i.e., $\hat{\vect{s}}=1/m \sum_{j = 1}^m \vect{h}_j$, which we then feed to the state-feedback controller. The control policy is learned using an off-policy actor-critic algorithm, Deep Deterministic Policy Gradient (DDPG)~\cite{lillicrap2016ContinuousControlDeep}, which is suitable for continuous state space and action space problems. This is achieved by continuous approximations of the policy and value functions by neural networks; the architecture is composed of an actor, i.e., the policy network, $\vect{\pi}(\vect{s}|\vect{\theta}^\pi)$, which outputs the deterministic action $\vect{a}$, and a critic, i.e., the action-value network, $\vect{Q}(\vect{s}, \vect{a}|\vect{\theta}^Q)$, which estimates the expected value of taking action $\vect{a}$ on the state $\vect{s}$ under the policy $\vect{\pi}$. The actor is trained to maximize the output of the critic, whereas the critic is trained to minimize the temporal difference error. As the agent interacts with the environment, we perform data assimilation on the observations and obtain the tuples containing $\big(\hat{\vect{s}}(t), \vect{a}(t), \hat{r}(t), \hat{\vect{s}}(t+1)\big)$, which are stored in an experience replay buffer, and sampled in batches for training in an off-policy fashion. 
\section{Results}\label{sec:results}
\begin{figure}
    \centering
    \includegraphics[width=\textwidth]{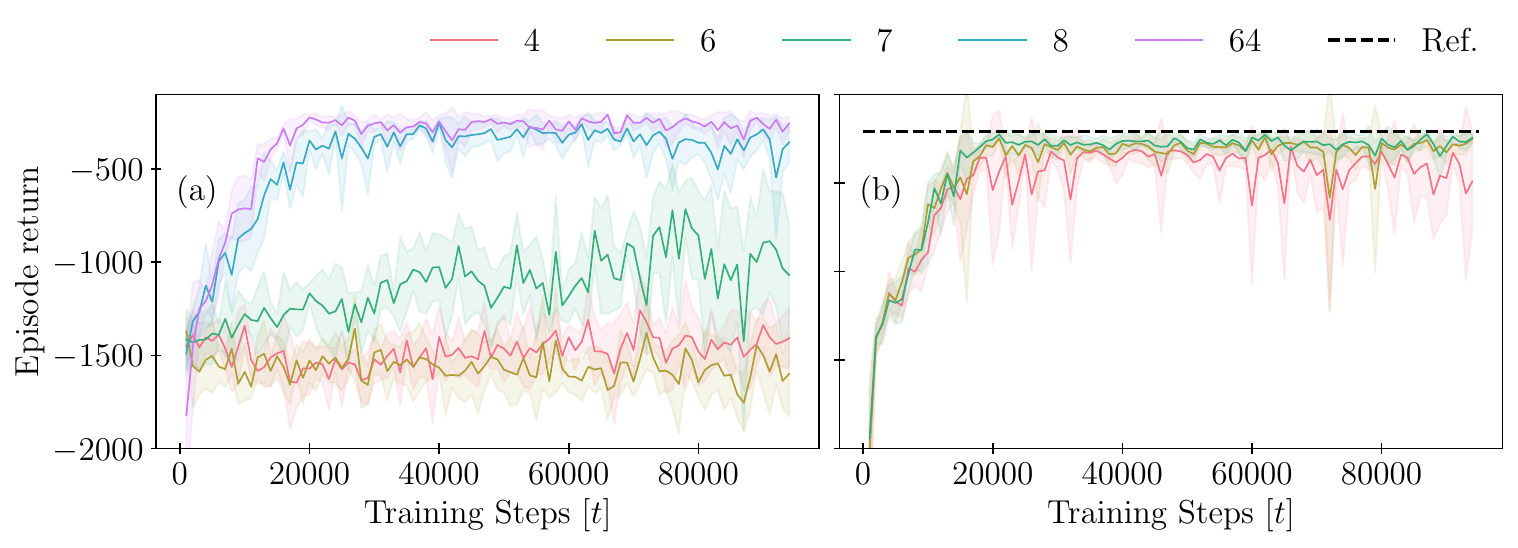}
    \caption{Training performance of (a) model-free and (b) DA-MBRL on the KS equation with different number of sensors. Shown non-discounted returns of training episodes across 5 runs (mean and 1 standard deviation). Ref. indicates the maximum return achieved by the model-free algorithm with full observability (i.e., $n_o = n_s = 64$).}
    \label{fig:return_comparison}
\end{figure}
\begin{figure}
    \centering
    \includegraphics[width=\textwidth]{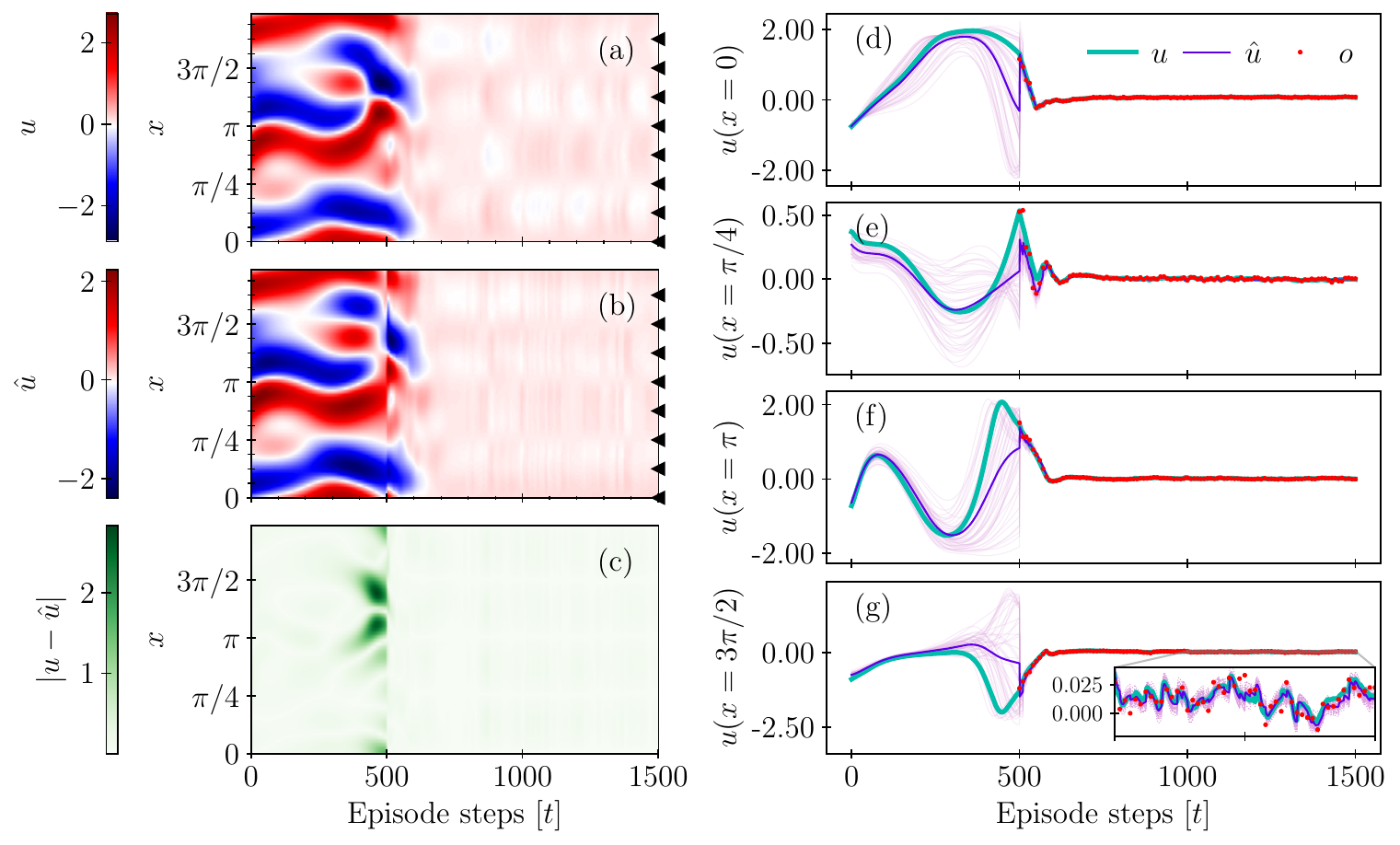}
    \caption{Evaluation episode of the DA-MBRL on the KS equation with 4 sensors. (a) The true state of the system, (b) the estimated state, (c) the reconstruction error, and (d-g) the observations and their ensemble estimates. The triangles in (a,b) indicate the actuator locations. The observations and control begins after 500 time steps.}
    \label{fig:eval_episode}
\end{figure}

We demonstrate our approach on the Kuramoto-Sivashinsky (KS) equation\footnote[1]{The code is available at \texttt{github.com/MagriLab/DA-RL}}, which is a 1D partial differential equation (PDE) that exhibits spatiotemporal chaos and has been employed as a benchmark for studying the control of chaotic and turbulent flows~e.g.,~\cite{gomes2017StabilizingNontrivialSolutions,bucci2019ControlChaoticSystems,weissenbacher2025ReinforcementLearningChaotic}
\begin{equation}
    \frac{\partial u}{\partial t} + \frac{\partial^2 u}{\partial x^2} + \nu \frac{\partial^4 u}{\partial x^4} + \frac{1}{2}\left(\frac{\partial u}{\partial x}\right)^2 = f,
\end{equation}
where $u(x, t)$ is the velocity on a $2\pi$-periodic spatial domain, i.e., $u(x, t) = u(x+2\pi, t)$, $\nu$ is the viscosity parameter, and $f$ is the forcing term to incorporate control. For $\nu < 1$ the trivial solution is linearly unstable and the system experiences bifurcations and chaotic regimes as $\nu$ is decreased further. In our study, we select $\nu = 0.08$, which results in a weakly chaotic regime often studied~e.g.,~\cite{bucci2019ControlChaoticSystems}. The control term is defined by a mixture of Gaussians
\begin{equation}
    f(x,t) = \sum_{i = 1}^{n_a}a_i(t) \exp\left(-\frac{(x-x_{a,i})^2}{2\sigma_a^2}\right),
\end{equation}
where $n_a$ is the number actuators, $\vect{a}(t) = [a_1(t); a_2(t); \dots; a_{n_a}(t)], \; a_{i}(t) \in [-1,1]$ are the mixture weights and the actions. The actuator locations are $\vect{x}_a = [x_{a,1}; x_{a,2}; \dots; x_{a,n_a}]$ and $\sigma_a$ is the width of each actuator. The PDE is discretized using a Fourier basis with $N = 64$ collocation points and solved using a third-order semi-implicit Runge-Kutta scheme~\cite{bucci2019ControlChaoticSystems} with a timestep of $\Delta t = 0.05$. The spatially-discretized state vector is $\vect{u}(t) = [u(0, t); u(2\pi/N, t); \dots; u(2\pi(N-1)/N, t)]$. We define the reward as a combination of the $\ell_2$-norms of the state and the action
\begin{equation}
    r(\vect{u}(t),\vect{a}(t)) = -\left(\frac{1}{\sqrt{N}}||\vect{u}(t+1)||_2 + \lambda||\vect{a}(t)||_2\right),
\end{equation}
which aims to stabilize the flow whilst using minimal actuation power weighted by $\lambda$. We first employ the model-free RL algorithm, i.e., DDPG, to obtain a baseline of the performance for different number of uniformly placed sensors measuring $u$. We initialize the system randomly and wait for a transient period of time for it to settle on the chaotic attractor, after which we start applying control. The training is performed in episodes of 1000 time steps. The evolution of the non-discounted episode returns are shown in Fig.~\ref{fig:return_comparison}(a) for varying number of sensors. Below $n_o = 8$ sensors, the model-free RL algorithm fails to discover a policy that stabilizes the flow. We test our proposed framework, DA-MBRL, on these challenging cases. First, a control-aware ESN with a reservoir state size of $n_h = 1000$ is trained on the full state data, i.e., $\vect{u}$, generated by applying random actuations to the system. Then during the training of RL, for each episode the ensemble is initialized around the attractor and then propagated for 1 Lyapunov time (approx. 500 time steps) to allow for sufficient divergence of the trajectories. We add zero-mean Gaussian noise with the covariance $\matr{C}_{\epsilon\epsilon}(t) = \left(0.1\max(\mathrm{abs}(\mathcal{M}(\vect{u}(t))))\right)^2\matr{I}$ on the observations and use these measurements to update the reservoir state every 10 time steps. The results of the DA-MBRL are shown in Fig.~\ref{fig:return_comparison}(b) for comparison, which demonstrate the effectiveness of the data assimilation step in making the controller effective in chaotic regimes with noisy and partial measurements. One evaluation episode after training is visualized in Fig.~\ref{fig:eval_episode} for the case of $n_o = 4$ sensors. The training and ESN model details for this case are provided in Tabs.~\ref{tab:training_params} and~\ref{tab:esn_params}, respectively.
\begin{table}[h!]
    \caption{Details and hyperparameters used for training the DA-MBRL agent for stabilisation of the KS equation.
    }\label{tab:training_params}
    \centering
    \begin{tabular}{l@{\hspace{2em}}l}
        Episode length & 1500 time steps \\
        Observation and control starts after & 500 time steps \\ 
        Number of random actuation steps & $5 \times 10^3$ (5 episodes)\\
        Number of training steps & $95 \times 10^3$ (95 episodes)\\ 
        Buffer capacity & $10^5$ \\ 
        Actor hidden units & [256, 256] \\ 
        Critic hidden units & [256, 256] \\ 
        Activation function & ReLU \\ 
        Actor learning rate & $3\cdot10^{-4}$ \\ 
        Critic learning rate & $3\cdot10^{-4}$ \\ 
        Discount factor ($\gamma$) & 0.99 \\ 
        Soft update rate ($\tau$) & 0.005 \\ 
        Batch size & 256 \\ 
        Exploration standard deviation & 0.1 \\ 
    \end{tabular}
\end{table}
\begin{table}
    \caption{Details and optimal hyperparameters of the control-aware Echo State Network with a reservoir size of $n_h = 1000$ and connectivity $n_{conn} = 3$ that learns the actuated dynamics of the KS equation.
    }\label{tab:esn_params}
    \centering
    \begin{tabular}{l@{\hspace{2em}}l}
    Washout length & 100 time steps \\
    Number of training steps & $56 \times 10^3$ (40 episodes) \\
    Leak rate ($\alpha$) & 0.23 \\
    Tikhonov coefficient ($\lambda$) & $10^{-6}$ \\
    Spectral radius ($\rho$) & 0.07 \\
    Input scaling ($\xi_{in}$) & 0.23 \\
    Action scaling ($\xi_{a}$) & 0.51 \\ 
    \end{tabular}
\end{table}
\section{Conclusions and Future Directions}\label{sec:conclusions}
Controlling partially-observed systems is challenging for traditional model-free reinforcement learning (RL) methods. We develop a framework that integrates (i) a predictive data-driven model (here, we use a control-aware Echo State Network), (ii) data assimilation for state estimation (here, we use an Ensemble Kalman Filter), and (iii) an off-policy actor-critic RL algorithm. We test the data-assimilated model-based RL (DA-MBRL) framework on the Kuramoto-Sivashinsky equation, which is a chaotic partial differential equation. We show that the model-free RL algorithm cannot discover a stabilizing policy below a certain number of sensors, i.e., under partial observability. The DA-MBRL successfully stabilizes the chaotic flow in these cases, opening up possibilities for the RL-based control of partially observed chaotic flows. Future work will test the framework on more complex, higher dimensional flow systems.
\par
$\textbf{Acknowledgements.}$ This research has received financial support from the UKRI AI for Net Zero grant EP/Y005619/1 and the ERC Starting Grant No. PhyCo 949388. The authors thank Georgios Rigas and Max Weissenbacher for their valuable insights and parts of the code.

\bibliographystyle{splncs04}
\bibliography{mybibliography}

\begin{thebibliography}{10}
\providecommand{\url}[1]{\texttt{#1}}
\providecommand{\urlprefix}{URL }
\providecommand{\doi}[1]{https://doi.org/#1}

\bibitem{brunton2015ClosedLoopTurbulenceControla}
Brunton, S.L., Noack, B.R.: Closed-{{Loop Turbulence Control}}: {{Progress}}
  and {{Challenges}}. Applied Mechanics Reviews  \textbf{67}(5),  050801 (Sep
  2015). \doi{10.1115/1.4031175}

\bibitem{bucci2019ControlChaoticSystems}
Bucci, M.A., Semeraro, O., Allauzen, A., Wisniewski, G., Cordier, L., Mathelin,
  L.: Control of chaotic systems by deep reinforcement learning. Proceedings of
  the Royal Society A: Mathematical, Physical and Engineering Sciences
  \textbf{475}(2231),  20190351 (Nov 2019). \doi{10.1098/rspa.2019.0351}

\bibitem{evensen2009DataAssimilationEnsemblea}
Evensen, G.: Data {{Assimilation}}: {{The Ensemble Kalman Filter}}. Springer
  Berlin Heidelberg, Berlin, Heidelberg (2009). \doi{10.1007/978-3-642-03711-5}

\bibitem{gomes2017StabilizingNontrivialSolutions}
Gomes, S.N., Papageorgiou, D.T., Pavliotis, G.A.: Stabilizing non-trivial
  solutions of the generalized {{Kuramoto}}--{{Sivashinsky}} equation using
  feedback and optimal control: {{Lighthill}}--{{Thwaites Prize}}. IMA Journal
  of Applied Mathematics  \textbf{82}(1),  158--194 (Feb 2017).
  \doi{10.1093/imamat/hxw011}

\bibitem{jaeger2004HarnessingNonlinearityPredicting}
Jaeger, H., Haas, H.: Harnessing {{Nonlinearity}}: {{Predicting Chaotic
  Systems}} and {{Saving Energy}} in {{Wireless Communication}}. Science
  \textbf{304}(5667),  78--80 (Apr 2004). \doi{10.1126/science.1091277}

\bibitem{krishnamurthy2016PartiallyObservedMarkov}
Krishnamurthy, V.: Partially {{Observed Markov Decision Processes}}: {{From
  Filtering}} to {{Controlled Sensing}}. Cambridge University Press, 1 edn.
  (Mar 2016). \doi{10.1017/CBO9781316471104}

\bibitem{lillicrap2016ContinuousControlDeep}
Lillicrap, T.P., Hunt, J.J., Pritzel, A., Heess, N., Erez, T., Tassa, Y.,
  Silver, D., Wierstra, D.: Continuous control with deep reinforcement
  learning. In: 4th {{International Conference}} on {{Learning
  Representations}}, \{\vphantom\}{{ICLR}}\vphantom\{\} 2016, {{San Juan}},
  {{Puerto Rico}}, {{May}} 2-4, 2016, {{Conference Track Proceedings}}. arXiv
  (2016). \doi{10.48550/ARXIV.1509.02971}

\bibitem{novoa2024InferringUnknownUnknowns}
N{\'o}voa, A., Racca, A., Magri, L.: Inferring unknown unknowns:
  {{Regularized}} bias-aware ensemble {{Kalman}} filter. Computer Methods in
  Applied Mechanics and Engineering  \textbf{418},  116502 (Jan 2024).
  \doi{10.1016/j.cma.2023.116502}

\bibitem{rabault2019ArtificialNeuralNetworks}
Rabault, J., Kuchta, M., Jensen, A., R{\'e}glade, U., Cerardi, N.: Artificial
  neural networks trained through deep reinforcement learning discover control
  strategies for active flow control. Journal of Fluid Mechanics  \textbf{865},
   281--302 (Apr 2019). \doi{10.1017/jfm.2019.62}

\bibitem{racca2021RobustOptimizationValidation}
Racca, A., Magri, L.: Robust {{Optimization}} and {{Validation}} of {{Echo
  State Networks}} for learning chaotic dynamics. Neural Networks
  \textbf{142},  252--268 (Oct 2021). \doi{10.1016/j.neunet.2021.05.004}

\bibitem{racca2023ControlawareEchoState}
Racca, A., Magri, L.: Control-aware echo state networks ({{Ca-ESN}}) for the
  suppression of extreme events (Aug 2023). \doi{10.48550/arXiv.2308.03095}

\bibitem{weissenbacher2025ReinforcementLearningChaotic}
Weissenbacher, M., Borovykh, A., Rigas, G.: Reinforcement {{Learning}} of
  {{Chaotic Systems Control}} in {{Partially Observable Environments}}. Flow,
  Turbulence and Combustion  (Jan 2025). \doi{10.1007/s10494-024-00632-5}

\bibitem{xia2024ActiveFlowControl}
Xia, C., Zhang, J., Kerrigan, E.C., Rigas, G.: Active flow control for bluff
  body drag reduction using reinforcement learning with partial measurements.
  Journal of Fluid Mechanics  \textbf{981}, ~A17 (Feb 2024).
  \doi{10.1017/jfm.2024.69}

\end{thebibliography}
\end{document}